# Photometry and Spectroscopy of a Deep Algol-like Minimum of WW Vulpeculae in 2016

**David Boyd**

*West Challow Observatory, OX12 9TX, UK; davidboyd@orion.me.uk*

**Abstract**   We report analysis of photometry and spectroscopy of a deep Algol-like minimum of the pre-main-sequence star WW Vul in July and August 2016. This revealed substantial reddening due to absorption by circumstellar material. After dereddening, our spectra of WW Vul were consistent with spectral type A3V throughout the event. Hα is normally in emission in WW Vul. During the minimum, Hα emission dropped by ~30% and FWHM of the Hα line reduced by ~15%.

### 1. Background on WW Vul

WW Vulpeculae was included in a list of RW Aurigae stars published by Hoffmeister in 1949. These are photometrically variable stars not assigned to any particular class. It was also included in the "Second Supplement to the Mount Wilson Catalogue" of stars with spectral types B and A whose objective-prism spectra have bright hydrogen emission (Merrill and Burwell 1949). Herbig (1960) compiled a list of 26 stars with spectral type Ae or Be associated with nebulosity and having emission lines in their spectra which he suggested were in the process of contracting onto the main sequence. These stars have come to be known as Herbig Ae/Be (or HAeBe) stars. Herbig's conjecture that HAeBe stars were young pre-main-sequence stars was confirmed by Strom *et al.* (1972), who showed that they are pre-main-sequence stars surrounded by optically thick circumstellar shells of dust which are possibly disc-like. Over the following years many authors discussed the nature of HAeBe stars and WW Vul in particular, including among others Grinin *et al.* (1996), Grinin *et al.* (2001), Hernandez *et al.* (2004), Mendigutia *et al.* (2011), and references therein. WW Vul is also described as an UXOR star, a term invented by Herbst *et al.* (1994) to classify young variable stars with masses in the approximate range 2 to 8 $M_\odot$ and strong Hα emission lines whose variability characteristics are similar to those of the prototype UX Ori.

A consensus has emerged that these stars are surrounded by a protoplanetary disc-like envelope seen almost edge-on containing optically thick clouds which orbit the star within the disc. Material accretes from the cloud onto the young star. Variable obscuration by these clouds causes irregular changes and occasional deep Algol-like minima in their light curves. According to this model the star's brightest state represents its intrinsic luminosity. The circumstellar material contains larger dust grains than occur in the interstellar medium and these strongly redden light emerging from the star. Hα emission lines in the spectrum of these stars originate not in the stellar photosphere but in the circumstellar gas and a stellar wind.

By comparison with spectra of standard stars, Merrill and Burwell (1949) gave the spectral type of WW Vul as A3e, Mora *et al.* (2001) classified its spectral type as A2IVe, and Hernandez *et al.* (2004) used an extensive range of spectral features to assign spectral type A3 with an uncertainty of ±2 subtypes. For the purpose of our analysis, we will take the spectral type of WW Vul to be A3V. Fitzgerald (1970) gave the intrinsic color index $(B–V)_0$ of an A3V star as 0.08, while Pecaut and Mamajek (2013) gave $(B-V)_0 = 0.10$. We will adopt 0.09. By analysis of spectral lines in comparison with synthetic spectra, Mora *et al.* (2004) estimated the photospheric effective temperature of WW Vul as 9000 K, Hernandez *et al.* (2004) gave 8710 K, and Pecaut and Mamajek (2013) gave the effective temperature of an A3V star as 8600 K. Hernandez *et al.* (2004) gave the mass of WW Vul as $2.9\,M_\odot$, while Mendigutia *et al.* (2011) gave $2.5\,M_\odot$.

### 2. Observations

Examining the long-term AAVSO light curve of WW Vul (AAVSO 2024) shows that it experiences long periods when its magnitude varies by less than 0.5 magnitude, interrupted by aperiodic Algol-like minima up to 2 magnitudes deep. Only twice in the last 20 years, in 2016 and 2019, has the V magnitude of WW Vul been repeatedly recorded fainter than 12. Here we analyze the deep minimum of WW Vul in July and August 2016 which reached V magnitude 12.12.

Photometric observations were made at West Challow Observatory in Oxfordshire in the UK with a 0.35-m Schmidt-Cassegrain telescope and Astrodon B and V photometric filters. Images were bias, dark, and flat corrected and instrumental magnitudes obtained by aperture photometry using the software AIP4WIN (Berry and Burnell 2005). A weighted ensemble of five nearby comparison stars was used whose B and V magnitudes and errors were obtained from AAVSO Variable Star Chart X27286KL (AAVSO 2016). Instrumental B and V magnitudes were transformed to the Johnson UBV photometric standard using the measured (B–V) color index and atmospheric airmass for each image. Times were recorded as Julian Date (JD).

B and V magnitudes and (B–V) color indices recorded before, during, and after the deep minimum in 2016 are listed in Table 1 and plotted in Figure 1. The (B–V) color index of WW Vul became redder as it faded before levelling out at minimum as shown in Figure 2. During the fade, the V-band flux of WW Vul dropped by ~75%. Recovery from minimum had an irregular profile, suggesting opacity of the material causing obscuration was changing from day to day.

Spectroscopic images were obtained concurrently with the photometry at the same location with a 0.28-m Schmidt-

Cassegrain telescope and a LISA slit spectrograph (Shelyak Instr. 2024) with spectral resolving power of ~1000. Images were bias, dark, and flat corrected, geometrically corrected, sky background subtracted, 1D spectrum extracted, and finally wavelength calibrated using the integrated argon-neon calibration source. Each WW Vul spectrum was corrected for instrumental and atmospheric losses by recording the spectrum of a nearby star from the MILES library of stellar spectra (Falcón-Barroso *et al.* 2011) observed at an airmass close to WW Vul. A list of recorded spectra is given in Table 2.

By recording photometry and spectroscopy concurrently, we were able to use the V magnitudes to calibrate all our spectra in absolute flux in units of erg/cm²/sec/Å using the method described in Boyd (2020). All photometry and spectra were submitted to, and are available from, the BAA Photometry Database and the BAA Spectroscopy Database (BAA 2024a, 2024b), respectively.

### 3. Circumstellar extinction and dereddening

Taking the intrinsic color index $(B–V)_0$ of WW Vul as 0.09, we found the color excess $E(B–V)$ for each of our spectra from $E(B–V) = (B–V) – (B–V)_0$. These $E(B–V)$ values are listed in Table 2. As selective extinction by circumstellar dust increased, it caused the star to fade, to redden, and $E(B–V)$ to increase.

Gaia DR3 (Bailer-Jones *et al.* 2021) gave the distance of WW Vul as $480 \pm 4$ pc. At this distance it is unlikely that the star will have experienced much galactic interstellar extinction. $R_V$ is the ratio of total visual extinction A(V) to selective extinction E(B–V). Hernandez *et al.* (2004) found that visual extinction measured for 39 HAeBe stars within 1 kpc including WW Vul is consistent with $R_V = 5.0$ rather than the usually adopted value of 3.1 for mean galactic interstellar extinction. They concluded that the high level of extinction for these stars is most likely produced by a combination of circumstellar material close to the star and material from their associated molecular clouds.

This circumstellar material reddens photospheric emission from WW Vul and changes the slope of the continuum. In our flux-calibrated spectra, WW Vul appeared to be closest to spectral type F2V from the Pickles atlas (Pickles 1998). To find its intrinsic spectral type we had to deredden our spectra. To do this we adopted $R_V = 5.0$ and used the relationships in Cardelli *et al.* (1989) to calculate $A(\lambda)/A(V)$ for each spectrum with the appropriate value of E(B–V) from Table 2. We applied this wavelength-dependent dereddening to each of our spectra. Figure 3 shows examples of our flux-calibrated spectra before and after dereddening and identifies some of the absorption lines in the spectra.

We compared each of our dereddened spectra of WW Vul with A2V, A3V, and A4V spectra from the Pickles atlas. These have similar resolution to our spectra. Of these, A3V gave the most consistent match to all our spectra, in agreement with published results. Figure 4 shows two dereddened spectra of WW Vul, one taken before the fade and one at minimum, together with Pickles A3V spectra.

Table 1. Log of B and V magnitudes and (B–V) color indices recorded before, during and after the deep minimum in 2016.

| Date | JD | B (mag) | V (mag) | (B–V) (mag) |
|---|---|---|---|---|
| 03-Jul-2016 | 2457573.4745 | 10.69 | 10.38 | 0.32 |
| 13-Jul-2016 | 2457583.4546 | 10.81 | 10.45 | 0.36 |
| 17-Jul-2016 | 2457587.5037 | 10.90 | 10.55 | 0.35 |
| 22-Jul-2016 | 2457592.4758 | 10.79 | 10.47 | 0.32 |
| 23-Jul-2016 | 2457593.4455 | 10.76 | 10.44 | 0.33 |
| 28-Jul-2016 | 2457598.4214 | 10.85 | 10.51 | 0.34 |
| 05-Aug-2016 | 2457606.4113 | 11.10 | 10.70 | 0.40 |
| 08-Aug-2016 | 2457609.4122 | 11.55 | 11.06 | 0.49 |
| 09-Aug-2016 | 2457610.3887 | 11.85 | 11.32 | 0.53 |
| 09-Aug-2016 | 2457610.4192 | 11.86 | 11.34 | 0.53 |
| 11-Aug-2016 | 2457612.4592 | 12.67 | 12.12 | 0.55 |
| 13-Aug-2016 | 2457614.4157 | 11.90 | 11.36 | 0.54 |
| 14-Aug-2016 | 2457615.3976 | 12.27 | 11.71 | 0.56 |
| 15-Aug-2016 | 2457616.4231 | 11.94 | 11.39 | 0.55 |
| 16-Aug-2016 | 2457617.4820 | 11.72 | 11.20 | 0.52 |
| 18-Aug-2016 | 2457619.3639 | 11.76 | 11.25 | 0.52 |
| 19-Aug-2016 | 2457620.3705 | 11.80 | 11.30 | 0.50 |
| 22-Aug-2016 | 2457623.3906 | 11.30 | 10.86 | 0.44 |
| 23-Aug-2016 | 2457624.3707 | 11.16 | 10.75 | 0.41 |
| 26-Aug-2016 | 2457627.3938 | 10.98 | 10.60 | 0.38 |
| 30-Aug-2016 | 2457631.3637 | 10.87 | 10.52 | 0.35 |
| 31-Aug-2016 | 2457632.4334 | 10.84 | 10.49 | 0.35 |
| 11-Sep-2016 | 2457643.3562 | 10.84 | 10.52 | 0.32 |
| 22-Sep-2016 | 2457654.3663 | 10.90 | 10.54 | 0.36 |

*Note: Average uncertainties in B are 0.014, in V are 0.012 and in B–V are 0.019.*

Table 2. Log of spectroscopic observations of WW Vul during the deep minimum in 2016 with color excess and integrated Hα emission line flux.

| Date | JD | E(B–V) | Hα emission line flux (erg/cm²/sec) |
|---|---|---|---|
| 17-Jul-2016 | 2457587.5037 | 0.26 | 2.13E–12 |
| 23-Jul-2016 | 2457593.4455 | 0.24 | 2.08E–12 |
| 05-Aug-2016 | 2457606.4113 | 0.31 | 1.48E–12 |
| 08-Aug-2016 | 2457609.4122 | 0.40 | 1.41E–12 |
| 09-Aug-2016 | 2457610.3887 | 0.44 | 1.35E–12 |
| 11-Aug-2016 | 2457612.4592 | 0.46 | 1.32E–12 |
| 14-Aug-2016 | 2457615.3976 | 0.47 | 1.34E–12 |
| 15-Aug-2016 | 2457616.4231 | 0.46 | 1.33E–12 |
| 16-Aug-2016 | 2457617.4820 | 0.43 | 1.28E–12 |
| 22-Aug-2016 | 2457623.3906 | 0.35 | 1.73E–12 |
| 26-Aug-2016 | 2457627.3938 | 0.29 | 1.75E–12 |
| 11-Sep-2016 | 2457643.3562 | 0.23 | 1.82E–12 |

*Note: The estimated uncertainty in line flux is 15%.*

### 4. Hα emission

Figures 3 and 4 show that our spectra of WW Vul have the broad Balmer absorption lines expected in a normal A3V star but with a narrower emission peak within the Hα absorption. This emission has been attributed to photospheric emission causing excitation in the circumstellar material and stellar wind. As the star faded, the Hα absorption dip weakened while the emission component remained strong. Figure 4 shows a small emission feature also appeared within the Hβ absorption line at minimum.

To measure absolute flux in the Hα emission line, we first scaled the Pickles A3V spectrum to match the flux level of the continuum adjacent to the Hα line in each of our dereddened WW Vul spectra and binned it to match the resolution of our spectra. We subtracted this from each of our WW Vul spectra to give the Hα emission flux in excess of photospheric emission. We then integrated this flux over the wavelengths of the line to give the total flux in the Hα line in that spectrum in erg/cm$^2$/sec. These data are listed in Table 2. The uncertainty in Hα flux is a combination of the uncertainties in determining the continuum level outside the line and in measuring the flux in the line.

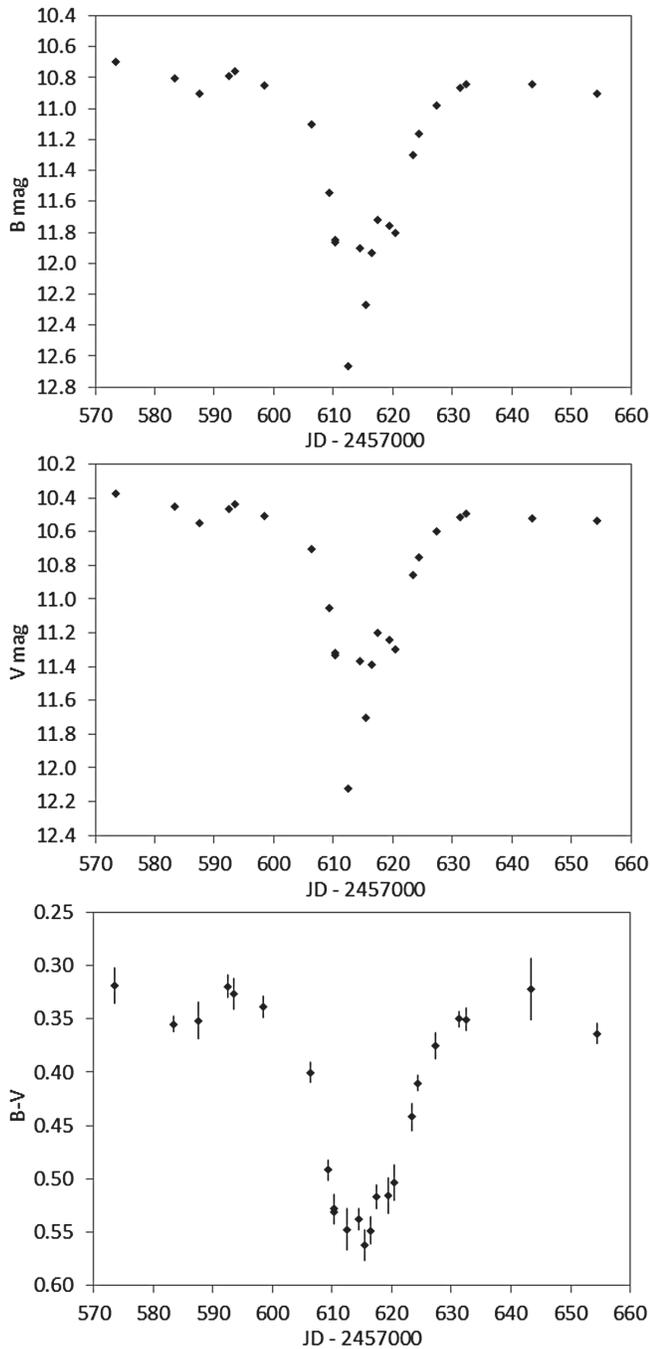

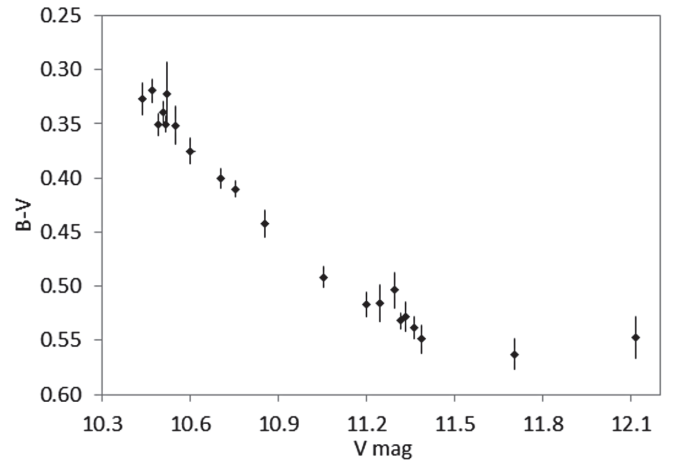

Figure 2. Variation of (B–V) color index as V magnitude faded.

Figure 1. Variation of B and V magnitudes and (B–V) color index before, during, and after the deep minimum of WW Vul in 2016.

Previous analyses of Hα emission (e.g. Mendigutia *et al.* 2011) have usually been in terms of equivalent width since, in general, the absolute flux in their spectra was not known.

By flux-calibrating our spectra, we were able to measure the strength of Hα emission in absolute flux. Figure 5 shows how flux in the continuum at Hα, Hα equivalent width, and Hα emission flux varied during the deep minimum in 2016. Hα equivalent width increased by a factor of three during the fade but because the continuum simultaneously decreased, Hα emission flux actually reduced by ~30% compared to its mean value before and after the event. This indicates that the source of Hα emission is partially blocked by the circumstellar material causing the fade.

Figure 6 shows the changing profile of the Hα emission line before, during, and after the fade. Before and after the fade the line was double-peaked, suggestive of a rotating circumstellar disc seen edge-on, while at minimum it became single-peaked. To investigate whether the width of the line varied during the fade, we fitted the emission peak with a Gaussian profile. Although the lines were sometimes double-peaked, a Gaussian profile fitted the rising and falling edges of all the lines well and so gave an accurate measurement of their full width at half maximum (FWHM). These FWHM measurements are plotted in Figure 7 and show that the Hα line became ~15% narrower during the fade. The Gaussian fits also gave the central wavelength of each line. These showed the lines were systematically displaced by 0.3 ± 0.3 Å towards the blue with respect to the rest wavelength of the line. Unfortunately there is no radial velocity for WW Vul available in Gaia DR3.

Figure 8 is a composite plot using color coding to illustrate the correlation between changes in spectral flux with wavelength and time in the upper panel and the variation in V magnitude with time in the left panel during the 2016 minimum of WW Vul. The red dots in the left panel mark the times of the spectra in the upper panel. The flux in the lower right panel is color-coded in ergs/cm$^2$/sec/Å according to the color scale on the right, with color linearly interpolated between the spectral profiles in the upper panel. This 2D color-coded plot provides a way of visualising the correlation between changes in spectral flux, V magnitude, and time that is not provided by separate plots of each parameter.

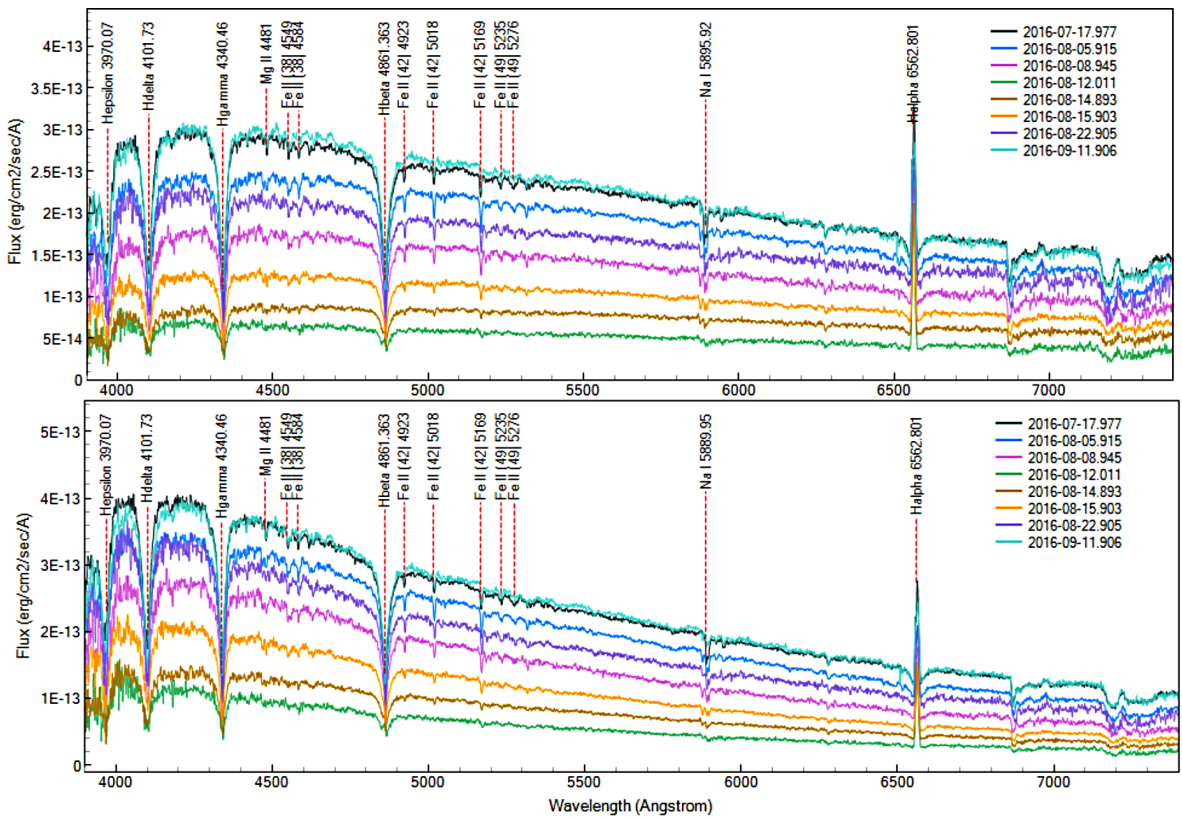

Figure 3. Examples of flux-calibrated spectra of WW Vul before (upper) and after (lower) dereddening, with identification of some of the absorption lines.

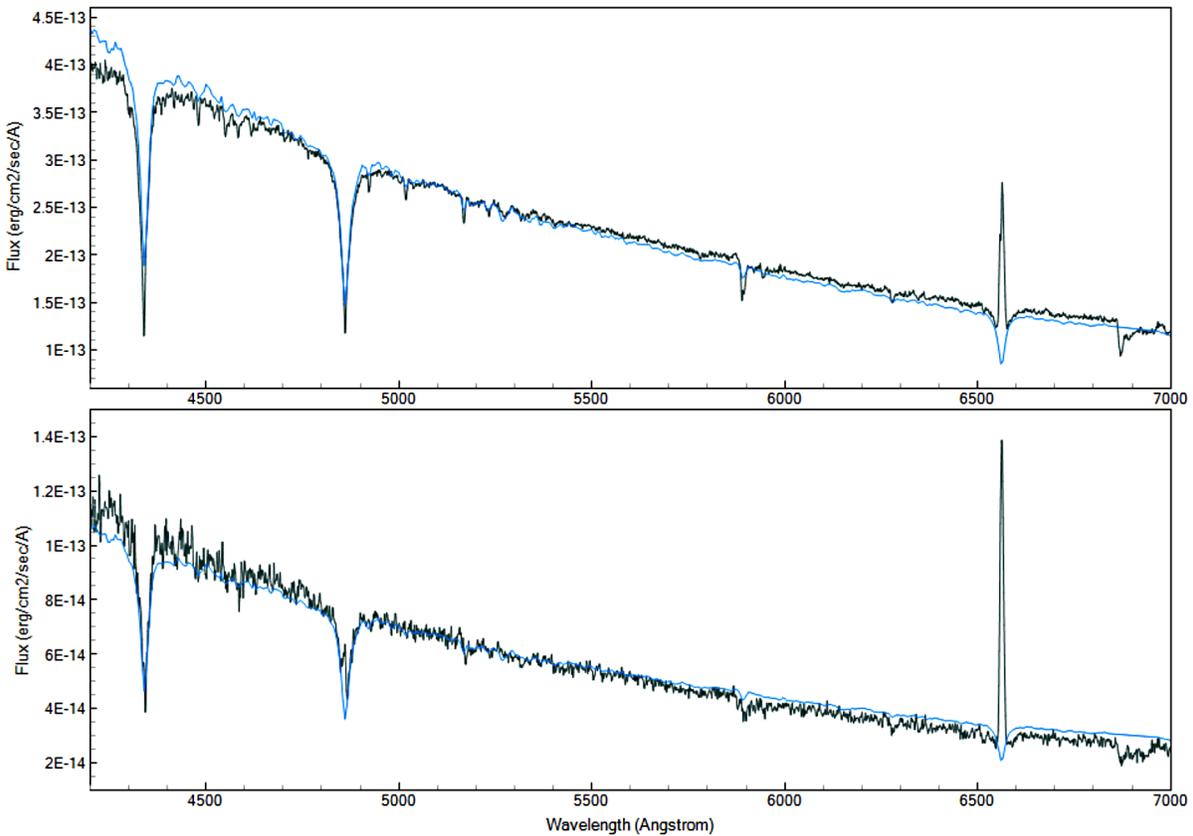

Figure 4. Comparing two dereddened spectra of WW Vul, one taken before the fade (upper) and one at minimum (lower), both with an A3V spectrum from the Pickles atlas (Pickles 1998).

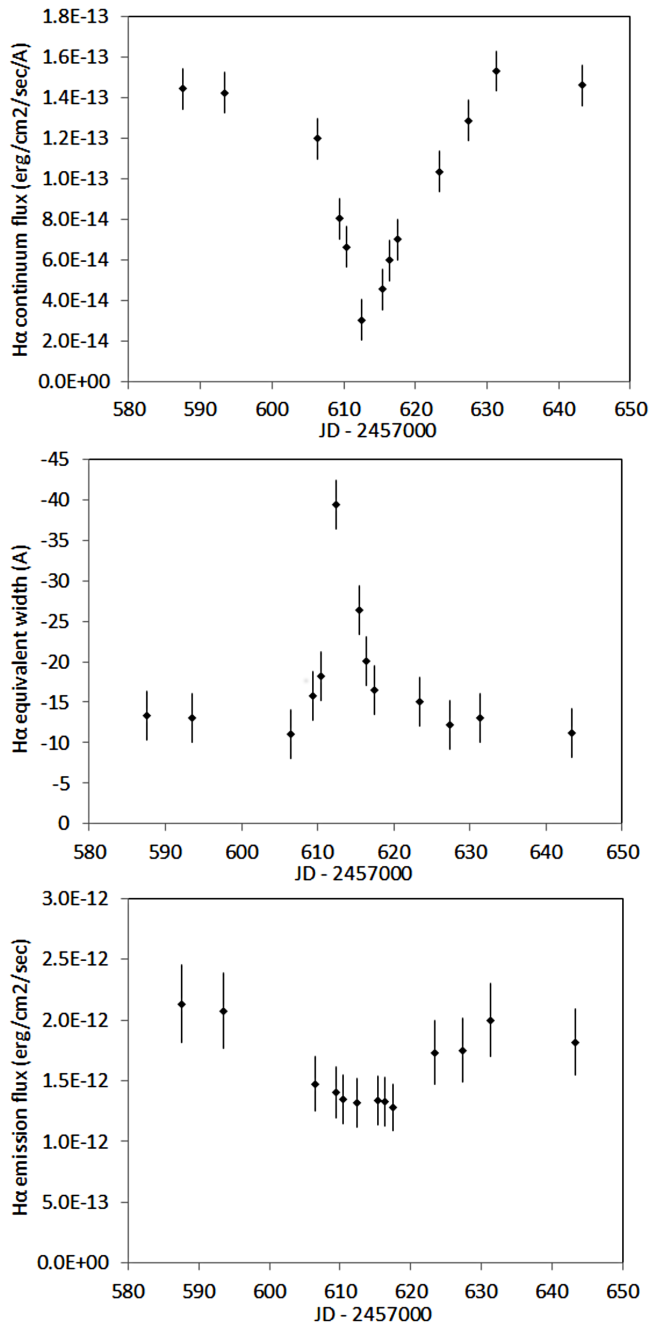

Figure 5. Continuum flux, equivalent width, and emission flux at Hα before, during, and after the deep minimum of WW Vul in 2016.

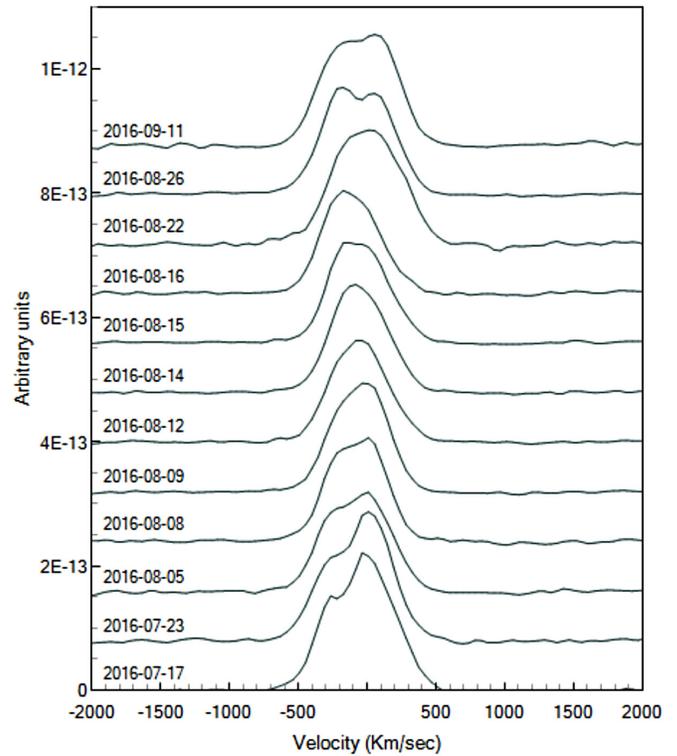

Figure 6. Hα emission line profile in spectra recorded during the 2016 minimum of WW Vul showing the line was double peaked before and after the fade but became single peaked at minimum.

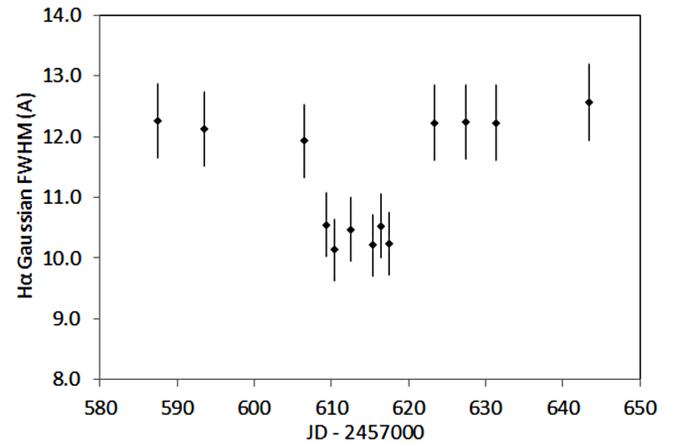

Figure 7. FWHM of Gaussian fits to Hα emission lines before, during, and after the 2016 minimum of WW Vul.

## 5. Summary

We observed a deep minimum of the pre-main-sequence star WW Vul in July and August 2016 with concurrent B- and V-band photometry and low resolution spectroscopy. We observed behavior consistent with the general consensus that transient obscuration by dense clouds of circumstellar dust in our line of sight substantially reduces direct photospheric emission and changes the slope of the continuum. We observed a ~75% drop in V-band flux of WW Vul and progressive reddening during the fade. We used our V-band photometry to flux calibrate our spectra and our measured values of E(B–V) to deredden our spectra. We found that the spectral type of WW Vul after dereddening consistently matched spectral type A3V throughout the fade. We measured the Hα emission lines and found their flux dropped by ~30% and their FWHM reduced by ~15% during the fade. The profile of the Hα line was double-peaked before and after the fade but became single-peaked at minimum.

## 6. Acknowledgements

We are grateful to the referee for a constructive review which helped us improve the paper. We also acknowledge the use of AAVSO Variable Star Charts and the work of developers of the Astropy package and other contributors to this valuable community resource.

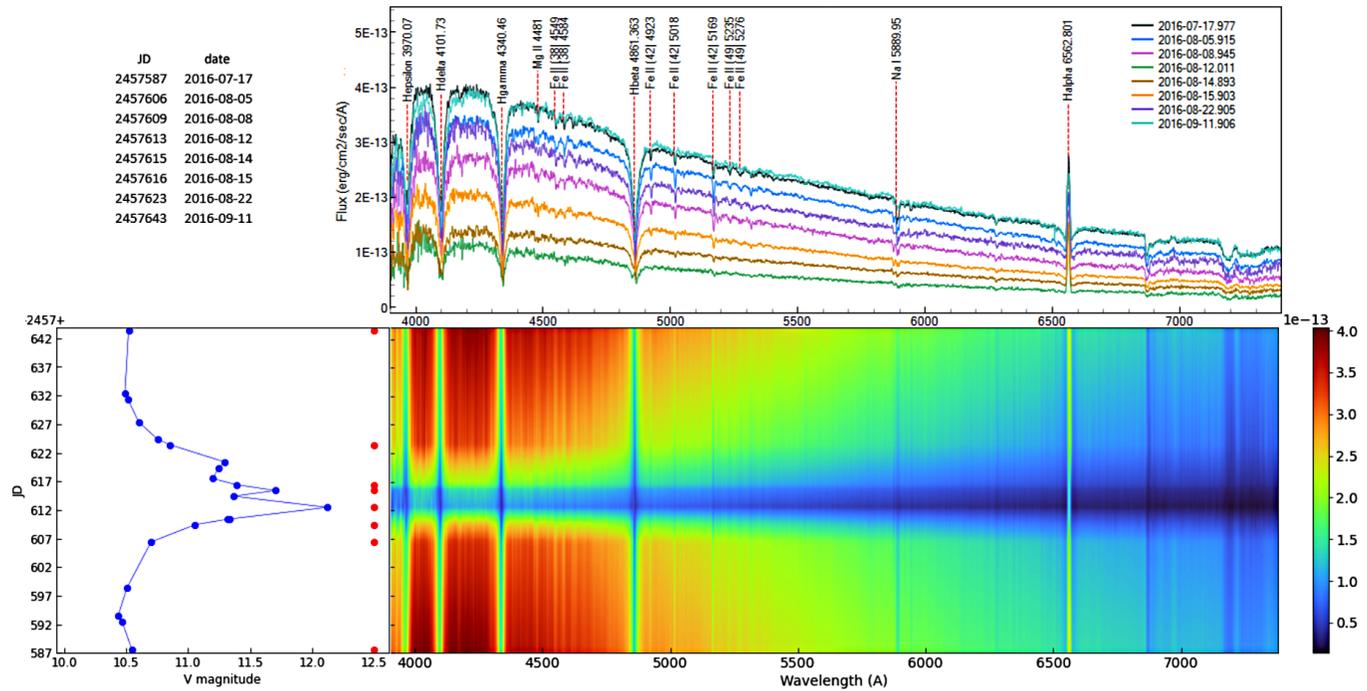

Figure 8. Composite plot using color coding to illustrate the correlation between changes in spectral flux with wavelength and time in the upper panel and the variation in V magnitude with time in the left panel during the 2016 minimum of WW Vul. The red dots in the left panel mark the times of the spectra in the upper panel. The flux in the lower right panel is color-coded in ergs/cm²/sec/Å according to the color scale on the right with color linearly interpolated between the spectral profiles in the upper panel.

---

[1] AAVSO (2024), https://www.aavso.org/LCGv2/index.htm?DateFormat=Julian&RequestedBands=&view=api.delim&ident=WW%20Vul&fromjd=460490&tojd=2460660&delimiter=@@@